%% file: main.tex
\documentclass[10pt,conference]{IEEEtran}
\IEEEoverridecommandlockouts
\usepackage{cite}
\usepackage{amsmath,amssymb,amsfonts}
\usepackage{algorithmic}
\usepackage{graphicx}
\usepackage{textcomp}
\usepackage{xcolor}
\usepackage{blindtext}
\usepackage{hyperref}
\usepackage{braket}
\usepackage{quantikz}
\usepackage{orcidlink}

\ifCLASSOPTIONcompsoc
\usepackage[caption=false,font=normalsize,labelfont=sf,textfont=sf]{subfig}
\else
\usepackage[caption=false,font=footnotesize]{subfi g}
\fi
\usepackage[subpreambles=true]{standalone}
\def\BibTeX{{\rm B\kern-.05em{\sc i\kern-.025em b}\kern-.08em
    T\kern-.1667em\lower.7ex\hbox{E}\kern-.125emX}}
    
\DeclareMathOperator*{\argmax}{arg\,max}
\DeclareMathOperator*{\argmin}{arg\,min}

\def\linkurl#1{\url{#1}}

\begin{document}

\title{Efficient QAOA Architecture for Solving Multi-Constrained Optimization Problems}

\makeatletter
\newcommand{\linebreakand}{%
  \end{@IEEEauthorhalign}
  \hfill\mbox{}\par
  \mbox{}\hfill\begin{@IEEEauthorhalign}
}
\makeatother
\author{%
\IEEEauthorblockN{%
David Bucher\IEEEauthorrefmark{1}\textsuperscript{$\orcidlink{0009-0002-0764-9606}$},
Daniel Porawski\IEEEauthorrefmark{1},
Maximilian Janetschek\IEEEauthorrefmark{1},
Jonas Stein\IEEEauthorrefmark{1}\IEEEauthorrefmark{2}\textsuperscript{$\orcidlink{0000-0001-5727-9151}$},
Corey O'Meara\IEEEauthorrefmark{3}\textsuperscript{$\orcidlink{0000-0001-7056-7545}$},
\\
Giorgio Cortiana\IEEEauthorrefmark{3}\textsuperscript{$\orcidlink{0000-0001-8745-5021}$},
and Claudia Linnhoff-Popien\IEEEauthorrefmark{2}\textsuperscript{$\orcidlink{0000-0001-6284-9286}$}}
\IEEEauthorblockA{\IEEEauthorrefmark{1}\textit{Aqarios GmbH, Munich, Germany}}
\IEEEauthorblockA{\IEEEauthorrefmark{2}\textit{LMU Munich, Institute for Computer Science, Munich, Germany}}
\IEEEauthorblockA{\IEEEauthorrefmark{3}\textit{E.ON Digital Technology, Hannover, Germany}}
\IEEEauthorblockA{\{david.bucher, daniel.porawski, maximilian.janetschek\}@aqarios.com}
}


\maketitle

\begin{abstract}
This paper proposes a novel combination of constraint encoding methods for the Quantum Approximate Optimization Ansatz (QAOA).
Real-world optimization problems typically consist of multiple types of constraints. To solve these optimization problems with quantum methods, normally, all constraints are added as quadratic penalty terms to the objective, which expands the search space and increases problem complexity.
This work proposes a general workflow that extracts and encodes specific constraints directly into the circuit of QAOA: One-hot constraints are enforced through $XY$-mixers that restrict the search space to the feasible sub-space naturally. Inequality constraints are implemented through oracle-based Indicator Functions (IF).
This paper focuses on the numerical benchmarks of the combined approach for solving the Multi-Knapsack (MKS) and the Prosumer Problem (PP), a modification of the MKS in the domain of electricity optimization.
To this end, we introduce computational techniques that efficiently simulate the two presented constraint architectures. Since $XY$-mixers restrict the search space, specific state vector entries are always zero and can be omitted from the simulation, saving valuable memory and computing resources.
We benchmark the combined method against the established QUBO formulation, yielding a better solution quality and probability of sampling the optimal solution. Despite more complex circuits, the time-to-solution is more than an order of magnitude faster compared to the baseline methods and exhibits more favorable scaling properties.
\end{abstract}


\input{sections/10_introduction}
\input{sections/20_background}
\input{sections/30_related_work}\input{sections/40_methods}
\input{sections/50_numerical_experiments}
\input{sections/60_conclusion}

\bibliographystyle{IEEEtranDoiTest}  
\bibliography{bstcontrol,main} 
\end{document}

%% file: sections/10_introduction.tex
\section{Introduction}\label{sec:introduction}

Many industry-relevant Combinatorial Optimization Problems (COPs), like the unit commitment problem~\cite{koretsky2021a} or the Prosumer Problem (PP)~\cite{mastroianni2024}, can be modeled as Multi-Knapsack (MKS) problems~\cite{awasthi2023}. Due to their NP-hardness, MKS and its base case, the single Knapsack Problem, pose a significant challenge for classical optimization algorithms~\cite{pisinger2005, awasthi2023}.

The recent advent of Quantum Computing (QC) technologies has sparked the application of QC to optimization~\cite{abbas2024}. Ever since Grover showed theoretical speedup for unstructured database search~\cite{grover1996}, the community began researching quantum algorithms that can solve optimization problems more efficiently than their classical counterparts, e.g., quantum minimum finding~\cite{durr1999}. The primary motivation stems from the fundamental property of quantum mechanical systems, \emph{superposition}, which allows qubits, the basic information units in QC, to be in $\ket{0}$ and $\ket{1}$ simultaneously. Consequently, many qubits can be in the superposition of the entire search space of a problem, including the solution itself. An optimization algorithm subsequently aims to boost the probability of measuring the solution bit-string.

Quantum optimization is primarily pursued with Quadratic Unconstrained Binary Optimization (QUBO) problems since they are isomorphic to Ising Hamiltonians with quadratic interactions, a fundamental physical many-body model~\cite{lucas2014a}. These models can be prepared in purpose-built machines, called Quantum Annealers (QAs), which use the adiabatic theorem of quantum mechanics to arrive at the optimal solution~\cite{farhi2000}.
For universal QC, the Quantum Approximate Optimization Ansatz (QAOA) has emerged as the most promising algorithm. It can be seen as a variational variant of the digitized simulation of the adiabatic process~\cite{farhi2014}.

Nevertheless, besides the maximum cut problem, relevant optimization problems, like MKS and PP, rely on constraints, which need to be encoded as quadratic penalties into the objective of a QUBO to be solved by QAOA~\cite{lucas2014a, glover2019}. However, solution quality usually drastically decreases due to the more complex energy landscape caused by the penalties and added slack variables~\cite{roch2023a, montanez-barrera2024a}. Therefore, constraint-preserving circuit variants for QAOA, like $XY$-Mixers for one-hot constraints, have emerged as encouraging alternatives to quadratic penalties~\cite{wang2020, hadfield2019a, fuchs2022, bartschi2020}. These circuit variants are a clear advantage of universal QC and QAOA compared to QA since they naturally reduce the search space, increasing overall performance.

This paper presents a combined architecture of QAOA that enforces one-hot constraints with $XY$-mixers and penalizes inequality constraints with a step function computed through a Quantum Phase Estimation (QPE) oracle, called Indictor Function (IF)~\cite{bucher2025}. Besides the work of Ref.~\cite{deller2023a}, which focuses on qudit (quantum digit) optimization, the literature has not investigated combinations of constraint-preserving architectures. We evaluate the performance and compare it to standard QUBO encoding using the MKS and PP instances. We utilize a specialized state vector simulator that employs several techniques to reduce the required state vector dimension. The paper is structured as follows: First, we will introduce the optimization problems and the QAOA in Sec.~\ref{sec:background}, followed by a brief discussion of the related work in Sec.~\ref{sec:related_work}. After Sec.~\ref{sec:methods} explains the circuit design and simulation techniques, Sec.~\ref{sec:experiments} presents the results of the numerical experiments before concluding with Sec.~\ref{sec:conclusion}.

%% file: sections/20_background.tex
\section{Background}\label{sec:background}
This section briefly introduces the COPs considered throughout the text and explains the QAOA algorithm. 

\subsection{Combinatorial Optimization Problems}

\subsubsection{QUBO}
The QUBO problem for $N$ binary variables is defined by an upper triangular matrix $Q\in \mathbb{R}^{N\times N}$ as finding the minimum to the following cost function
\begin{align} 
    \argmin_{x\in \{0,1\}^N} \sum_{i\leq j} Q_{i,j} x_i x_j,
\end{align}
over a discrete search space $x \in \{0,1\}^N$.
This optimization problem is isomorphic to the quadratic Ising Hamiltonian by the transformation $x_i \rightarrow \frac{1}{2}(1 - Z_i)$, where $Z_i$ is the Pauli-$Z$ operator acting on qubit $i$. The ground state of that physical system encodes the solution of the QUBO.
Many COPs have already been formulated as QUBO to be solved with QC~\cite{lucas2014a, glover2019}. These transformations require reformulating any constraint $c^Tx = b$ as a penalty $(c^Tx-b)^2$ added to the QUBO objective. Inequality constraints are first transformed into equality constraints by introducing a slack variable that covers the feasible differences between the left and right sides of the constraint.

\subsubsection{Multi-Knapsack Problem}

The objective of the MKS is to assign a set of $n$ items to $m$ Knapsacks, such that the overall value contained in the Knapsacks is maximized. Each item has a defined value $v_i$ and Knapsack-dependent weight $w_{i,j}$ and can be assigned to at most one Knapsack~\cite{hess2024}. Every Knapsack has a total weight capacity of $W_j$, giving us the expression
\begin{gather}
    \argmax_{x\in\{0,1\}^{n\times m}} \sum_i v_i \sum_j x_{i,j}  \quad \text{s.t.} \quad 
    \sum_j x_{i,j} \leq  1 \quad \forall i=1,\dots,n\nonumber\\
    \text{and}\quad  \sum_i w_{i,j}x_{i,j} \leq W_j \quad\forall j = 1,\dots,m .
    \label{eq:mks-orig}
\end{gather} 

\subsubsection{Prosumer Problem}
The PP is an application related to operating the future electricity grid. Our formulation is derived from Ref.~\cite{mastroianni2024}. The optimization problem runs over a discrete time horizon $m$ with $n$ loads having pre-defined load profiles $\ell$ running for $\tau$ time steps. The goal is to minimize the money spent on electricity, which is determined by a time-variable electricity price rate of $r$, through scheduling the starting time of the required loads, i.e.,
\begin{alignat}{2}
\argmin_{x \in \mathcal{D}} &\sum_i \sum_{t=0}^{m-\tau_i} x_{i,t} \sum_{t'=0}^{\tau_i-1} r_{t+t'}\ell_{i,t'}.
    \\ \text{s.t.}\quad 
    &\sum_t^{m-\tau_i} x_{i,t} = 1 \quad &&\forall i = 1,\dots,n
    \nonumber\\
    \text{and}\quad
    &\sum_i \sum_{t'=T_-(t, \tau_i)}^{T_+(t, \tau_i)} \ell_{i,t'-t+\tau_i - 1}x_{i,t'} \leq W \,\,\,&&\forall t = 1,\dots,m,\nonumber
    \label{eq:pp}
\end{alignat}
where $T_-(\tau, t) = \max\{0, t-\tau_i+1\}$ and $T_+(\tau, t) =\min\{t, m-\tau_i\}$.
A feasible solution requires that each load be started precisely once and that the total capacity limit per timestep never be surpassed. The problem is related to MKS, but there is additional cross-influence between Knapsack placements.

\subsection{Quantum Approximate Optimization Ansatz}

The QAOA algorithm was first introduced to solve the maximum cut optimization problem by Farhi et al.~\cite{farhi2014}. However, QAOA can be employed at its core to find the ground state of any Ising cost Hamiltonian $H_C$, isomorphic to QUBO or higher-order optimization~\cite{glos2022}. 
The algorithm employs a variational approach that alternates between two Hamiltonians: the cost Hamiltonian $H_f$ and a mixing Hamiltonian $H_M$, which is typically the $X$-Mixer $H_M=-\sum_i \sigma_x$. For a QAOA circuit of depth $p$, the algorithm prepares the trial state:
\begin{align}
    \ket{\beta, \gamma} = e^{-i\beta_pH_M} e^{-i\gamma_p H_f} \cdots e^{-i\beta_1H_M} e^{-i\gamma_1 H_f} \ket{+}^{\otimes N},
\end{align}
where $\gamma = (\gamma_1, \dots, \gamma_p)$ and $\beta = (\beta_1, \dots, \beta_p)$ are variational parameters and $\ket{+}^{\otimes N}$ is the initial state with all $N$ qubits in equal superposition, i.e., the ground state of $H_M$. 

The objective is to optimize these $2p$ parameters to minimize the expectation value:
\begin{align}
    C(\beta, \gamma) = \bra{\beta,\gamma} H_C \ket{\beta, \gamma}
\end{align}
Since the optimal solution is the lowest energy state, this procedure amplifies the probability of sampling a low-energy solution at the end of the algorithm. A classical optimizer performs the outer optimization loop by updating the variational parameters, making QAOA a hybrid quantum-classical approach. As $p$ increases, the solution quality generally increases, with the theoretical guarantee that QAOA converges to the optimal solution in the limit $p \rightarrow \infty$~\cite{farhi2014}. However, recent developments have also shown that fixed parameters (increasing $\gamma$'s and decreasing $\beta$'s) with finite $p$ can lead to sufficient quality solutions, such that the classical optimization overhead can be omitted~\cite{sack2021a, montanez-barrera2024}.

%% file: sections/30_related_work.tex
\section{Related Work}\label{sec:related_work}

The MKS problem has been studied several times in the context of quantum optimization~\cite{hess2024, awasthi2023, guney2025, sharma2024}. Ref.~\cite{sharma2024} focused on a novel, qubit-efficient quantum random access optimization algorithm, while the other works mainly investigated QAOA and QA on various unconstrained problem reformulations.
By comparing different QAOA variants with QA, Ref.~\cite{awasthi2023} showed that warm-starting QAOA, meaning initializing the algorithm with a biased initial state, performs best alongside QA. In our work, we focus on quantum algorithmic enhancements of QAOA. Thus, we deliberately exclude warm starting to avoid distorting the results. Ref.~\cite{hess2024} mainly studied different quadratic penalty encodings of the Knapsack inequality constraint into QAOA. As a result, they found that not including slack variables is generally favorable but approximates the optimization. If no approximation can be tolerated, it is best to integrate the slack variables into the circuit but omit them in the measurements.

The original work~\cite{mastroianni2024}, and inspiration of our PP problem formulation, examined the performance of the recursive QAOA algorithm~\cite{bravyi2022} towards solving PP transformed into QUBO. They successfully demonstrated good results even on real hardware. Nevertheless, the heuristic nature of the recursive algorithm could also be solely responsible for the solution quality in this case.

Since our work focuses explicitly on a mixture of constraint-preserving QAOA architectures, we introduce related constraint-preserving techniques here. At its core, to enforce a constraint in QAOA, we need to start in a superposition of only feasible states and construct a mixer operator that only causes amplitude transfer in the feasible sub-space. If the mixer satisfies the conditions defined in Ref.~\cite{hadfield2019a}, the modified QAOA efficiently searches the feasible sub-space only. While conceptually simple, the implementation of general constraints is complex. One-hot constraints are simple constraints with a known structure of the feasible subspace. Namely, exactly one binary variable can be set to 1, i.e. $\sum_i x_i = 1$. For them, $XY$-mixers have been developed (further detailed in Sec.~\ref{sec:methods}) and extensively studied in Refs.~\cite{cook2020a, wang2020, bucher2024} with the result that they are relatively efficient in implementation cost and lead to a considerable performance benefit. Constraint-preserving mixer methods have been further investigated and generalized to a wide selection of constraints in Refs.~\cite{hadfield2019a, hadfield2021, fuchs2022, bartschi2020}.

However, non-structured constraints, like the general inequality constraint, are significantly more challenging to implement. While various ansätze exist, e.g., based on Zeno dynamics or tree generators~\cite{herman2023, wilkening2024,christiansen2024}, they often struggle with obstructive circuit implementation requirements. As an alternative, cost function methods based on quantum oracles for constraint satisfaction have emerged for these kinds of constraints~\cite{grandrive2019, kea2023, bucher2025}. While, again, they are searching an extended space including infeasible states, they have proven more efficient than the quadratic penalties known from QUBO transformations~\cite{kea2023, bucher2025}. We rely on the penalty-based IF-QAOA proposed in Ref.~\cite{bucher2025}.

%% file: sections/40_methods.tex
\section{Methods}\label{sec:methods}

This section discusses our methods for generating and simulating problem-specific QAOA circuits for the MKS and PP. We first unify the problems into a format consisting only of inequality and one-hot constraints. Then, we apply $XY$-mixers and IF-QAOA for the constraints to generate custom QAOA circuits. Finally, we will introduce an efficient simulation algorithm, leveraging the reduced search space, helping us to investigate problem sizes well above the limits of current qubit state vector simulators.

\subsection{General Structure of Constrained Optimization Problems} 

The problems at our focus consist of linear objectives with $Q$ one-hot and $m$ linear inequality constraints, formalized as follows
\begin{align}
    \argmin_{x\in\{0, 1\}^N} f(x)  \quad \text{s.t.} \quad g_j(x) &\geq 0 \quad \forall j = 1,\dots,m\nonumber\\
    \sum_{i \in D} x_i &= 1 \quad \forall D\in \{D_1, \dots,D_Q\},
    \label{eq:general}
\end{align}
where $f$ and $g_j$ are considered linear functions defined over the binary variables. Consequently, the exact upper and lower bounds, $g^+_j$ and $g^-_j$, can be computed. Each $D_i$ contains $d_i > 2$ elements\footnote{A $d_i = 2$ one-hot constraint can be encoded by a single qubit and $d_i = 1$ is forced to be one all the time.}, required to follow that $D_i \cap D_j = \emptyset\ \forall i\neq j $, meaning one-hot constraints do not overlap since that would be impossible to encode with $XY$-mixers. If this is the case, we can alleviate the issue by removing a subset of the one-hot constraints through transforming them into quadratic penalties.

Furthermore, we assume here that the image of every $g_j$ is resolvable by integers, i.e., $g_j(\{0,1\}^N) \subset \mathbb{Z}$, to construct a fully-resolved IF, as required by Ref.~\cite{bucher2025}.

The PP, as defined in Eq.~\eqref{eq:pp}, is already in the required form\footnote{The transformation of the inequality constraints to $g_j(x) \geq 0$ is trivial.} of Eq.~\eqref{eq:general} and satisfies the stated conditions.
Yet, MKS~\eqref{eq:mks-orig} has no one-hot constraint, but an item can be placed in \emph{at most} one Knapsack (set-packing constraint). This property can be exploited by introducing a \emph{dummy} Knapsack, denoted by the binary variable $y$, with no weight limit. It serves as a container for all items not selected in the main Knapsacks, yielding the one-hot constrained formulation of Eq.~\eqref{eq:mks-orig}
\begin{gather}
    \argmax_{x, y} \sum_i v_i \sum_j x_{i,j}  \,\,\, \text{s.t.} \,\,\,
    \sum_j x_{i,j} + y_i = 1 \,\,\, \forall i=1,\dots,n\nonumber\\
    \text{and}\quad  \sum_i w_{i,j}x_{i,j} \leq W_j \quad\forall j = 1,\dots,m .
    \label{eq:mks-reformulated}
\end{gather}


\subsection{Circuit Generation}
Based on the general constrained optimization format, we can generate a customized QAOA circuit with a combined constraint-preserving architecture using the steps presented below. The generator pipeline is designed to be modular, such that certain steps can be enabled or disabled. The circuit will contain $N = K+\sum_i d_i$ compute qubits, where $K$ is the number of binary variables not associated with a one-hot constraint.

\subsubsection{One-hot constraints to XY-Mixers}

\begin{figure*}
    \subfloat[IF+XY Circuit Diagram\label{fig:circ}]{\includegraphics[scale=0.75]{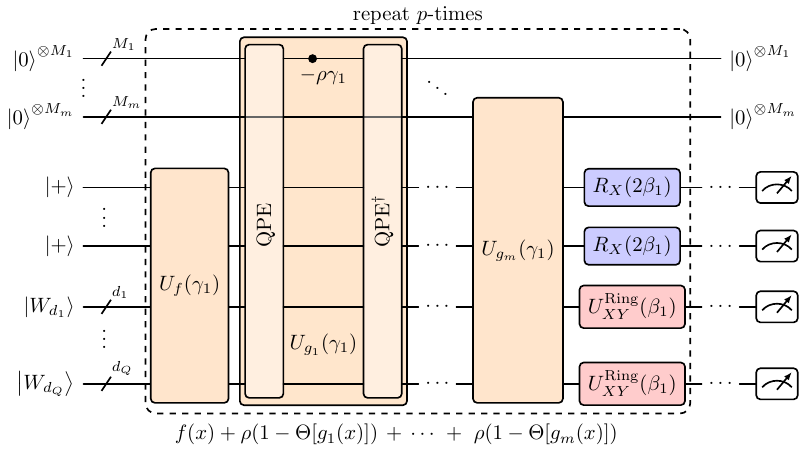}}
    \hfill
    \subfloat[Simulation Loop\label{fig:sim}]{\scalebox{0.8}{\input{drawings/sv_tensor}}}
    \caption{Panel (a) shows the circuit diagram of the combined architecture, implementing both $XY$-mixers for one-hot constraints and IF penalties for inequality constraint satisfaction. Panel (b) depicts the simulation loop for a mixture of qubits and qudits. The mixer layer is split into two sections: First, the $R_X$ mixer is applied on all qubits by reshaping the state tensor into a matrix and applying batched gate multiplication. Afterwards, the qudit legs are expanded and the $U_{XY}^\text{Ring}$ matrices are contracted onto the legs. Both operations rely on NVIDIA's \texttt{cuQuantum} SDK.} 
    \label{fig:circandsim}
\end{figure*}

Following Ref.~\cite{hadfield2019a}, we need two quantum routines to restrict QAOA to the one-hot constraint subspace effectively: First, a state preparation circuit generates the equal superposition of all feasible states, which is defined as
\begin{align}
    \ket{W_d} = \frac{1}{\sqrt{d}}(\ket{10\cdots 0} + \ket{010\cdots 0}+ \cdots +\ket{0\cdots01}),
\end{align}
being the equal superposition of all feasible (Hamming weight 1) bit-strings of size $d$ associated with a single one-hot constraint. The employed circuit from Ref.~\cite{cruz2019} constructively generates $\ket{W_d}$ by applying parametrized SWAP gates ($\cos(\phi) I +\sin(\phi) \mathrm{SWAP}$) in a tree structure, leading to a logarithmic depth implementation.

Together with the other variables, the combined initial state is a product state of $\ket{W}$ states and $\ket{+}$ states, i.e.,
\begin{align}
    \ket{\psi_0} = \ket{+}^{\otimes K} \otimes \left[\bigotimes_i \ket{W_{d_i}}\right],
\end{align}
which is depicted on the left-hand side of Fig.~\ref{fig:circ}.

The second part is the $XY$-mixer, which is formally defined as the evolution of complete $XY$-Hamiltonian~\cite{wang2020,hadfield2019a},
\begin{align}
    H_{\text{XY}}(D) =\frac{1}{2}\sum_{i\in D_i } \sum_{j \in D_j} (X_iX_j + Y_j Y_i),
\end{align}
where $X$ and $Y$ denote the respective Pauli operators. $\ket{W_d}$ is the ground state of $H_\text{XY}$.
However, due to the fully connected Hamiltonian and the non-commuting $XY$-terms when sharing a qubit, implementation requires Trotterization of the evolution, leading to lengthy circuits that grow linearly with the one-hot size $d$~\cite{wang2020}.

Yet, since $XX+YY = 2(\ket{01}\bra{10} + \ket{10}\bra{01})$,
\begin{align}
U_{XY}(\beta)  &= e^{-i(XX +YY)\beta / 2} = R_{YY}(\beta)R_{XX}(\beta)
\end{align}
already fulfills the property of only transitioning between feasible states, thereby satisfying the conditions from Ref.~\cite{hadfield2019a}, we can construct a mixer operation out of $U_{XY}$ routines. A low-cost implementation of a mixer that provides probability transfer between all states given sufficient QAOA-depth $p$ is the \emph{ring mixer}, which only places $U_{XY}$'s between neighboring qubits in a brick-wall configuration. For instance, when $d=5$, the ring mixer is constructed through~\cite{hadfield2019a}
\begin{align}\label{eq:ring-mixer}
    \scalebox{0.8}{
    \begin{quantikz}[row sep=4pt, column sep=6pt, rounded corners=4pt]
    & \gate[2, style={inner sep=0}] {U_{XY}} & \qw                                    &\gate[5,style={fill=none}, label style={yshift=1.15cm}]{U_{XY}} & \qw \\
    & &                                        \gate[2, style={inner sep=0}] {U_{XY}} &\linethrough & \qw \\
    & \gate[2, style={inner sep=0}] {U_{XY}} & &                                       \linethrough & \qw \\
    & &                                        \gate[2, style={inner sep=0}] {U_{XY}} &\linethrough & \qw \\
    & &                                        &                                        & \qw
    \end{quantikz}
    }
    =U_\text{last} U_\text{odd}   U_\text{even} = U_{XY}^\text{Ring}.
\end{align}
Here, $U_\text{last}$ is only required because the number of qubits is odd. In the even case, this final unitary is discarded. The $U_{XY}^\text{Ring}$ mixer is placed on all selected one-hot constrained qubits, as shown in Fig.~\ref{fig:circ}.

\subsubsection{Inequality Constraints to Indicator Functions}

Knapsack constraints can be integrated into QAOA through non-linear step-function penalties added to the objective function using IF~\cite{bucher2025}. With the Heaviside Theta defined as $\Theta(z) = 1$ if $z \geq 0$ and $0$ otherwise, this can be formalized as follows
\begin{multline}
    \argmin_x f(x) \quad \text{s.t.}\quad g_j(x)\geq 0 \quad\forall j=1,\dots, m\\ \xrightarrow{\quad} \argmin_x \underbrace{f(x) + \rho \sum_j (1-\Theta[g_j(x)] )}_{\eqcolon \tilde{f}(x)}.
\end{multline}
The penalty parameter $\rho$ has to be chosen such that no infeasible state has a lower modified objective value $\tilde{f}(x)$ than the best feasible solution to the original problem. To fulfill this, $\rho$ needs to be greater or equal to the difference between the best feasible solution $x^*$ and the best solution that violates a single constraint, denoted as $\tilde{x}^*$, i.e., $\rho \geq f(x^*) - f(\tilde{x}^*) $. Determining both solutions exactly is infeasible, so we approximate the values with respective bounds. A valid upper bound for $f(x^*)$ is given by any feasible solution $x^1$, which can generally be found through heuristics. A valid lower bound for $f(\tilde{x}^*)$ is computed by solving the relaxed problem formulation $f^{\text{relaxed}}$ with $0 \leq x \leq 1$ without inequality constraints. Consequently, we set $\rho  = f(x^1) - \min_x f^{\text{relaxed}}(x)$.

The quantum circuit implementation of the IF consists of two QPEs for each constraining function $g_j$ and an ancillary register, as detailed in Ref.~\cite{bucher2025}. The register size is $M_j = 1 + \lceil \log_2 \max\{|g^-|, |g^+|\} \rceil$. The initial QPE encodes constraint satisfaction information in the first ancillary qubit, i.e., whether $g_j(x)$ is $\geq 0$ or not. Depending on this state, we apply a phase gate that adds a constant penalty if $g_j(x)$ is unsatisfied, summarized as follows
\begin{gather}\label{eq:if-operations}
    \ket{x}\ket{0} \xrightarrow{\text{QPE}_{g_j}} \ket{x}\ket{g_j(x)<0} \xrightarrow[\text{\small on ancilla qubit}]{P(-\rho\gamma)}\\ \{\ket{x}\ket{0}, e^{-i\rho \gamma} \ket{x}\ket{1}\} \xrightarrow{\text{QPE}^\dagger_{g_j}} e^{-i\rho\gamma(1 - \Theta[g_j(x)])} \ket{x}\ket{0}\nonumber.
\end{gather}
Since the IF application of each constraint reverts the compute register to $\ket{0}$, we may reuse this ancillary register for sequential application, requiring only one ancillary register for multiple constraints. However, since constraints typically depend on only a selection of binary variables, cf.\ MKS, we can apply the oracles for each constraint simultaneously on different registers, leading to shorter circuits and faster runtime at the expense of more utilized qubits. The entire circuit architecture is depicted in Fig.~\ref{fig:circ}, with a $U_{g_j}(\gamma)$ for each Knapsack constraint containing the subroutines of Eq.~\eqref{eq:if-operations}.

The constraint implementation via the IF has the following advantages: First, we have a bounded penalty magnitude that does not distort the energy spectrum as much as a quadratic penalty, the default QUBO approach uses~\cite{mirkarimi2024, bucher2025}. Second, even though the IF-method uses additional qubits to compute the constraint satisfaction, these qubits are not part of the optimization problem compared to the slack variable approach.

\subsubsection{Modular Architecture}

Since the pipeline is designed to be modular, we would like to compare different configurations, e.g., only using $XY$-mixers, against each other. This requires a final step in the workflow that integrates all remaining constraints that have not been handled by the methods above via default QUBO methods. Namely, adding them into the objective via quadratic penalty functions and adding slack variables to span the feasible space of the inequality constraints~\cite{lucas2014a}. Essentially, the model is transformed into the following QUBO, when not using any of the constraint-preserving components discussed above:
\begin{multline}
    \argmin_{x,y} f(x) + \eta \sum_D \left(1 - \sum_{i\in D}x_i \right)^2 \\+ \eta\sum_j \left( g_j(x) - \sum_{\ell}2^\ell y_{j,\ell} + a_j y_{j,M_j} \right)^2,
\end{multline}
where each inequality constraint $j$ has $M_j = \left\lceil\log_2g_j^+\right\rceil$ binary slack variables $y_{j\ell}$~\cite{lucas2014a}. The last coefficient for $y_{j,M_j}$ is given by $a_j = g^+_j - 2^{M_j-1}+1$, allowing the slack variable to cover the integer range from $0$ to the upper bound $g_j^+$. Consequently, any value of $g_j(x)$ smaller than 0 will be penalized. We utilize D-Wave's software package \texttt{dimod}\footnote{\url{https://github.com/dwavesystems/dimod}} for the automatic translation of the constraints to penalty terms.

The modular pipeline allows us to compare four methods against each other \textbf{QUBO} (only quadratic penalty terms), \textbf{XY} (only $XY$-mixers, no IF), \textbf{IF} (only IF, no $XY$-mixers), and \textbf{IF+XY} (both IF and $XY$-mixers active). 

\subsection{Simulator}\label{sec:sim}

Next, we describe our simulation technique for the extended QAOA, using several enhancements to accelerate computation and enable problem sizes not accessible with ordinary state vector simulation.

\subsubsection{Data Layout}
Since only the feasible states w.r.t.\ the one-hot constraints have non-zero amplitude in the $XY$-mixer case, we can omit non-feasible states from the computation, leaving us with $d$ non-zero state vector entries for a one-hot constraint over $d$ binary variables. Effectively, this state can be seen as a $d$-level \emph{qudit} (quantum digit), i.e. $\ket{W_d} = \frac{1}{\sqrt{d}}(\ket{1}+\ket{2} +\dots + \ket{d})$~\cite{deller2023a}. Therefore, from now on, we will refer to any joint qubit system under one-hot constraint as qudit.

The state vector of the simulator can be imagined as a tensor with $K$ dimension-two legs for the qubits and a dimension-$d_i$ leg for every qudit, as depicted on the left-hand side of Fig.~\ref{fig:sim}. Formally, this state can be expressed as
\begin{align*}
    \ket{\psi} = \sum_{\substack{i_1,\dots,i_K \in \{0,1\}\\l_1=1,\dots,d_1\\l_Q=1,\dots,d_Q} }
    \psi_{i_1,\dots,i_K,l_1,\dots,l_Q}\ket{i_1,\dots,i_K,l_1,\dots,l_Q}.
\end{align*}
The number of the state tensor entries, and therefore the memory requirements, is $2^K \prod_i d_i$, which is drastically less than the full qubit representation $2^{K + \sum_i d_i}.$

\begin{figure}
\hrule\vspace{0.2em}
\begin{algorithmic}
\REQUIRE $\text{Term set } \mathcal{T}, \text{Starting bits }S, \text{Tensor shape}$
\STATE $f \gets \text{real, zero tensor of shape } (2,\dots,2,d_1,\dots,d_Q)$
\FORALL{$I = (i_1,\dots,i_N,k_1,\dots,k_Q)$}
\STATE $x \gets 0$ \hfill \textit{Initialize the key for current entry}
\FORALL{$i \in I, s \in S$}
\STATE $x \gets x\mathbin{|}(i\mathbin{\ll}s)$ \hfill \textit{Shift index to key start}
\ENDFOR
\FORALL{$(k, m, v) \in \mathcal{T}$}
\IF{$k = m \mathbin{\&}x$} 
\STATE $f_I \gets f_I + v$ \hfill \textit{Term key equals masked current key}
\ENDIF
\ENDFOR 
\ENDFOR
\RETURN $f$
\end{algorithmic}
\hrule
\caption{Algorithm for efficient brute-forcing based on bit keys and masks. Symbols $\mathbin{\&}$, $\mathbin{|}$ and $\mathbin{\ll}$ refer to bitwise operations.}
\label{fig:algo}
\end{figure}
\subsubsection{Precomputation through Brute-Forcing} Every QAOA layer applies the cost function for each basis state to the phase of the respective state vector entry. Since the solution can be read from the phase angle in the state vector entry, this is equivalent to brute-forcing. QAOA technically repeats this brute-forcing at every iteration. Therefore, there is a performance advantage in precomputing the cost function before simulation, as previous research has shown~\cite{lykov2023a, golden2023, stein2024b}. We extend the method described in Ref.~\cite{stein2024b} to accommodate the qudits in the brute-forcing algorithm, as detailed in Fig.~\ref{fig:algo}.

The algorithm iterates over every possible configuration and sums up the contributing terms using only a few bitwise operations.
The state of each qubit can be represented by a single bit, whilst the state of each qudit can be described by $\lceil \log_2d_i\rceil$ bits, requiring $K+\sum_i \lceil\log_2d_i\rceil$ bits to represent any qubit-qudit configuration. This means a 64-bit integer practically suffices to encode all state tensor entries. The starting positions $S$ store the location of the bits in the integer for one particular qubit or qudit. The polynomial cost function can then be defined as the set of terms, where each term consists of a \emph{key}, telling which configuration is required, a \emph{mask}, indicating which qubits and qudits are part of the term, and a \emph{value}, which is gained if the bits are in required configuration. Please refer to Ref.~\cite{stein2024b} for a more detailed algorithm description.

\paragraph*{Example} The cost function $f(w, x, y, z) = wx+ 2wy + 3wz -w$ needs to be brute-forced, s.t.\ $x +y+ z = 1$.
The first bit encodes $w$, and the second and third bits encode the qudit representation for $q = (x,y,z)$. The starting positions for the two variables are $s=0$ for $w$ and $s=1$ for $q$. The mask for the first three terms is $111$ because both $q$ and $w$ are present; for the last one, it is $001$. Finally, the term set can be defined as {\small $\mathcal{T}=\{(001,111,1), (011,111,2), (101,111,3), (001,001,-1)\}$} and brute-forcing can be executed, leading to $f=(0, 0, 0, 0, 1, 2)$.

Following Ref.~\cite{bucher2025}, the ancillary registers for the IF \emph{do not need to be simulated} since the classical step function can be evaluated within the brute-forcing. Each constraint $g_j(x)$ adds a penalty $\rho$ to $f(x)$ when unsatisfied. Thus, we can brute-force all $g_j$'s to arrive at the combined brute-forced $\tilde{f} = f + \rho \sum_i (1 - \Theta(g))$ IF cost function. This precomputed vector can then be applied to the QAOA state tensor, as depicted in Fig.~\ref{fig:sim}.

\subsubsection{State Tensor Size}
\begin{figure}
    \centering
    \includegraphics[width=\linewidth]{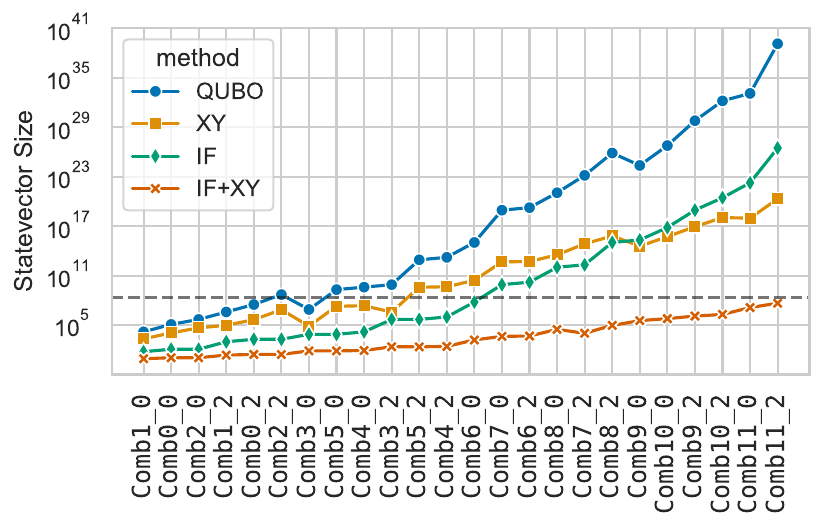}
    \caption{State vector size required for simulation depending on the PP instance and encoding method. The dashed line indicates the 28-qubit equivalent, which marks the maximum for our numerical simulations.}
    \label{fig:sv_size}
\end{figure}

Fig.~\ref{fig:sv_size} shows the number of state tensor entries for the four different methods regarding the PP instances, defined in Tab.~\ref{tab:problem-instances}. IF+XY has the most modest memory requirements of only $\prod_i d_i$, remaining below our simulation threshold of a 28 qubit equivalent, i.e. 4\,GB memory. The IF method raises to $2^{\sum_i d_i}$, XY to $2^M\prod_i d_i$, where $M = \sum_j M_j$ is the number of ancillary variables required to encode all inequality constraints. Finally, QUBO requires $2^{M+\sum_i d_i}$ entries, clearly being the most expensive for simulation.

\subsubsection{Simulation Schema}
The whole simulation procedure (Fig.~\ref{fig:sim}) starts by brute-forcing $\tilde{f}$ and initializing the state tensor in an equal superposition. Then, the first QAOA layer begins with applying $e^{-i\tilde{f}\gamma}$ to the state vector, followed by the mixer, being split into the qubit and qudit parts in our case. 

We utilize NVIDIA's \texttt{cuQuantum} SDK~\cite{bayraktar2023} for both parts. The regular $X$-mixer on the qubits is applied with \texttt{custatevec.apply\_matrix\_batched} using the broadcasted $R_X(2\beta)$ matrix on the reshaped state tensor. The reshaping turns the tensor into a matrix with all qubits in one dimension and qudits in the other, ready for batched gate application. By clever initial data layout, the entries are already in the required order in memory, so no computational overhead through reshaping is introduced.

For the qudits, we utilize to \texttt{cutensornet.contract} to contract the $d_i$-dimensional square $\hat U^\text{Ring}_{XY}(\beta)$ matrices onto the open qudit legs of the tensor (Fig.~\ref{fig:sim}), which is the sub-matrix of the $U^\text{Ring}_{XY}(\beta)$ in the feasible subspace. Since $U_{XY}(\beta)$ matrices, acting on the feasible subspace is exactly $R_X(2\beta)$, we can use $R_X$ to construct $\hat U^\text{Ring}_{XY}$, without projecting up to the $2^{d_i}$ space.

The ring mixer layers are block diagonal matrices in the feasible subspace, i.e.
\begin{align}
    \hat{U}_\text{even}(\beta) &= \mathrm{blockdiag}(R_X, R_X,\dots,1)\nonumber\\
    \hat{U}_\text{odd}(\beta) &= \mathrm{blockdiag}(1, R_X, \dots, R_X,)\\
    \hat{U}_\text{last}(\beta) &= \begin{pmatrix}(R_X)_{22} && (R_X)_{21}\\&I_{d_i-2}\\(R_X)_{12}&&(R_X)_{11}\end{pmatrix}\nonumber,
\end{align}
where the last mixer is special because it transitions between state $1$ and $d_i$. Note that in the case where $d_i$ is even, $U_\text{last}$ is not required, but the outer structure is transferred to $U_\text{odd}$, and $\hat{U}_\text{even} =\mathrm{blockdiag}(R_X,\dots,R_X) $.
Finally, we can multiply the three matrices like in Eq.~\eqref{eq:ring-mixer} to arrive at $\hat{U}^\text{Ring}_{XY}$, continue with the contraction, and ultimately start with the next QAOA layer.

%% file: drawings/sv_tensor.tex
\begin{tikzpicture}
\draw[thick, gray!50, ->] (4, 0.3) -- (4, -0.3);
\draw[thick, gray!50, ->] (3.2, -1.7) to[out=-135,in=-45] (0.5, -2);
\draw[thick, gray!50, ->] (0.5, 2) to[out=45,in=135]  (3.2, 1.7);
\node[rounded corners, fill=myellow, draw, minimum height=3.1cm] at (0.1cm, 0.1cm) {$\ket{\psi}$};
\node[rounded corners, fill=mblue, draw, minimum height=3.1cm] (tensor) at (0, 0) {$\ket{\psi}$};

\node[rotate=90] at (-1, 0) {Apply $\exp(-i\tilde{f}\gamma)$};

\draw[thick,dblue] (tensor.east) ++(0,1.3) -- ++(0.5,0) node[right] (l1)  {2};
\draw[thick,dblue] (tensor.east) ++(0,0.5) -- ++(0.5,0) node[right]{2};
\draw[very thick,dred] (tensor.east) -- ++(0.5,0) node[right]{$d_1$};
\draw[very thick,dred] (tensor.east) ++(0,-0.5) -- ++(0.5,0) node[right](l2){$d_2$};
\draw[very thick,dred] (tensor.east) ++(0,-1.3) -- ++(0.5,0) node[right]{$d_Q$};

\node[below, inner ysep=1pt] at (l1)  {\small \vdots};
\node[below, inner ysep=0pt] at (l2)  {\small \vdots};

\node[fill=white, fill opacity=0.5,text opacity=1] (dsv) at (4.8, 1.9) {\texttt{custatevec.apply\_matrix\_batched}};

\node[circle, fill=mblue, draw] (tensor2) at (4,1) {$\ket{\psi}$};
\draw[line width=3pt,dred] (tensor2.west) -- ++(-0.5, 0) node[left]{$\prod_i d_i$};
\draw[line width=3pt,dblue] (tensor2.east) -- ++(0.5, 0) node[right] (x) {$2^N$};

\draw[->] (dsv.south) ++(2,0) to[out=-90,in=0] node[fill=white, inner sep=1pt]{$R_X$} (x);

\node[circle, fill=mblue, draw] (tensor3) at (4,-1) {$\ket{\psi}$};
\draw[line width=3pt,dblue] (tensor3.west) -- ++(-0.5, 0) node[left]{$2^N$};

\node[circle, fill=mred, draw, inner sep=0] (u1) at (5.2, -0.0) {\tiny $U_{XY}$};
\node[circle, fill=mred, draw, inner sep=0] (u2) at (5.2, -0.75) {\tiny $U_{XY}$};
\node[circle, fill=mred, draw, inner sep=0] (u3) at (5.2, -1.8) {\tiny $U_{XY}$};

\node[above, inner ysep=10pt] at (u3) {\tiny \vdots};

\draw[very thick,dred] (tensor3.east) to[out=0, in=180] (u2.west);
\draw[very thick,dred] (tensor3.north east) to[out=45, in=180] (u1.west);
\draw[very thick,dred] (tensor3.south east) to[out=-45, in=180] (u3.west);

\draw[very thick,dred] (u1.east) -- ++(0.3, 0);
\draw[very thick,dred] (u2.east) -- ++(0.3, 0);
\draw[very thick,dred] (u3.east) -- ++(0.3, 0);

\node at (5.2, -2.4) {\texttt{cutensornet.contract}};

\end{tikzpicture}

%% file: sections/50_numerical_experiments.tex
\section{Numerical Experiements}\label{sec:experiments}

This section presents the numerical experiments performed for the MKS and PP problem instances using the simulation methods described above. Before investigating the results, we discuss how the experiments were conducted and define figures of merit for performance comparison. All experiments were performed on a NVIDIA RTX 4090 GPU.

\subsection{Experimental Setup and Evaluation Metrics}

The evaluation and optimization use the same cost function for every method (QUBO, IF, XY, and IF+XY). Namely, we start with $\tilde{f}$, but also add penalties for one-hot constraints
\begin{multline}\label{eq:cost-func-eval}
    F(x) = f(x) + \eta \sum_j (1 - \Theta(g_j(x))) \\
    +\eta\sum_{D} (1-\delta_{1, \sum_i x_i}),
\end{multline}
where $\delta_{i,j}$, is 1 if $i=j$ and 0 otherwise. Consequently, a constant penalty $\eta$ is added for every unsatisfied constraint, making all pipelines comparable\footnote{The second penalty is always zero for both XY variants.}. Furthermore, as only the binary variables of the original problem are required for this cost, we disregard any samples of the ancillary register, improving the performance of the slack variable approaches~\cite{hess2024}.

Optimization is done with the BFGS optimizer provided by \texttt{scipy} using a maximum of 100 iterations~\cite{scipy}. Furthermore, we use the interpolation recipe given by Ref.~\cite{zhou2020a}. This means we start with QAOA layers $p=1$, optimize the parameters, then interpolate to $p=2$, optimize again, and continue that pattern until we have reached $p=12$. With this method, the initial parameters for each optimization iteration are already satisfactory, leading to faster convergence and a general trend of increasing solution quality with $p$.

\subsubsection*{Metrics}
We evaluate optimization performance using three key metrics~\cite{bucher2024}. First, the solution quality is assessed through the \emph{Random-Adjusted} Approximation Ratio (RAAR)~\cite{bucher2025}, defined as follows
\begin{align}
    \text{RAAR} = \frac{\langle F(x)\rangle - \bra{\psi} F(x)\ket{\psi}}{\langle F(x)\rangle -f(x^*)}.
\end{align}
Here, $\langle F(x)\rangle$ represents the random average of the cost function from Eq.~\eqref{eq:cost-func-eval}. An RAAR value of 0 indicates performance equivalent to random sampling, while a value of 1 signifies consistent sampling of the perfect solution.

The second figure of merit is the probability of measuring the optimal solution at the end of the QAOA circuit, which can be obtained by taking the square of the respective state tensor entry\footnote{Not denoted here, but degenerate perfect solutions are aggregated.} $P^* = \left|\braket{x^*|\psi}\right|^2$. With slack variables in place, we trace over the slack qubits and evaluate the probability only in the initial problem register, similar to Ref.~\cite{hess2024}. Besides perfect solutions, one might also be interested in the likelihood of sampling solutions with a certain quality measure, e.g., at least 90\% of the optimal value. Formally, $P_{90} = \sum_{x\in \mathcal{D}_{90}} \left|\braket{x|\psi}\right|^2$, with $\mathcal{D}_{90} = \{x \in \mathcal{D} | f(x) \leq 0.9 f(x^*)\}$~\cite{hess2024}.

The third metric, Time-to-Solution (TTS), is estimated from $P^*$ as the runtime in terms of circuit layers required to measure the optimal solution once with 99\% confidence~\cite{bucher2024,albash2018a}. TTS has the advantage of also considering circuit complexity alongside sampling efficacy since it is defined as the expected number of shots times the number of circuit layers $\mathbf{L}(p)$ in the depth-$p$ QAOA:
\begin{align}\label{eq:tts}
    \text{TTS}_p  = \mathbf{L}(p)\left\lceil\frac{\log\,0.01 }{\log(1-P^*)}\right\rceil.
\end{align}
From TTS$_p$, we defined the optimal TTS$^*$
\begin{align}
    \text{TTS}^*  = \min_p \text{TTS}_p
\end{align}
which represents the minimum TTS over all tested $p$.

\subsubsection*{Circuit Layer Operations}

Generally, the QAOA circuit layers are defined as $\mathbf{L}(p) = \mathbf{L}_\text{init} + p(\mathbf{L}_\text{cost} + \mathbf{L}_\text{mixer})$. For the circuit layer counts, we assume all-to-all connectivity and two-qubit Pauli rotations being native gates.
To compute $\mathbf{L}(p)$ for the various pipelines, we investigate the different parts separately, starting with ordinary qubits, $\mathbf{L}_\text{init} = \mathbf{L}_\text{mixer} = 1$. The cost function layer $\mathbf{L}_f$ is given by the number of interactions in the cost function $f(x)$. The circuit layers may differ drastically depending on the structure of $f$, e.g., when penalties are enforced. To estimate the number of layers, we use the greedy edge coloring heuristic provided by \texttt{rustworkx}~\cite{treinish2022} on the interaction graph, as the chromatic number gives the smallest number of non-overlapping gate layers. When no IF is applied, $\mathbf{L}_\text{cost} = \mathbf{L}_f$.

The largest qudit $\max_i d_i = d_\text{max}$ is the bottleneck for the layers in the $\ket{W}$ state preparation, which is given by $\mathbf{L}_\text{init} = 2 \lceil\log_2 d_\text{max}\rceil$, following Ref.~\cite{cruz2019}. The ring mixer is itself not size-dependent, but whether $U_\text{last}$ is required. Since $R_{XX}, R_{YY}$ are considered single depth, the mixer layers are given by $\mathbf{L}_\text{mixer} = 4 \,(+2)$.

Since all IF penalties are evaluated in parallel, we first compute the layers of the phase part $\mathbf{L}_\text{phase}$ in the parallel QPEs via the chromatic number of the interaction graph, similar to $\mathbf{L}_f$. The interaction graph is constructed by adding an edge between every QPE qubit and the respective binary variables of the constraint. The remaining circuit consists of two QFTs and a single phase gate, which is again dominated by the biggest QPE register $\max_j M_j = M_\text{max}$. Total IF layers are given by $\mathbf{L}_\text{IF} = 2\mathbf{L}_\text{phase} + 2 (2 M_\text{max} - 1) + 1$~\cite{bucher2025}. The final QAOA layer cost is then the appropriate selection of all the presented parts.

\subsection{Multi-Knapsack Problem}
\begin{figure}
    \centering
    \includegraphics[width=\linewidth]{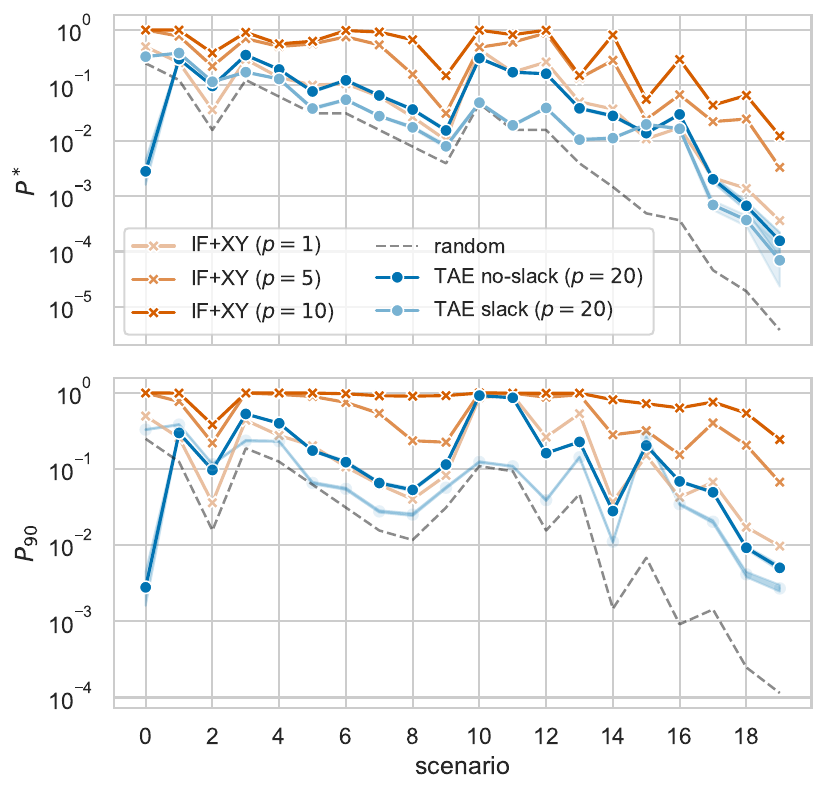}\\[-1em]
    \caption{$P^*$ and $P_{90}$ of the combined IF+XY (at $p=5,10$) in comparison to the most promising TAE (slack and no-slack at $p=20$) approach from Ref.~\cite{hess2024}. Multi-Knapsack instances start at scenario number 10. All smaller instance numbers are single Knapsack instances (no one-hot constraints and $XY$-Mixer necessary).}
    \label{fig:mks-results}
    \vspace{-1em}
\end{figure}

Hess et al.~\cite{hess2024} investigated $P^*$ and $P_{90}$ metrics for different constraint encodings when solving the MKS problem (Eq.~\eqref{eq:mks-orig}) with QAOA and variants. Their results show that the Trotterized Adiabatic Evolution (TAE)---QAOA without an optimization loop---achieves comparable performance to QAOA. They examined two well-performing constraint encodings: one using a QUBO formulation without slack variables but with an approximative equality penalty, and another incorporating slack variables while evaluating only the compute register, similar to Eq.~\eqref{eq:cost-func-eval}. Unlike our QUBO formulation, which requires one-hot constraints, their QUBO formulation does not require an additional dummy qubit for every item. Fig.~\ref{fig:mks-results} presents results for both TAE variants at $p=20$, taken from Ref.~\cite{hess2024}. TAE employs a sinusoidal schedule to predetermine  $\gamma_i =\delta s(i\delta)$, $\beta_i = \delta (1 - s(i\delta))$, with
$s(i\delta) = \sin^2\left[\frac{\pi}{2}\sin^2\left(\frac{\pi i}{2p}\right)\right]$
and $\delta=0.75$~\cite{hess2024}.

We benchmark these TAE results against our IF+XY QAOA approach at $p=1,5,10$ using the reformulation in Eq.~\eqref{eq:mks-reformulated} on the identical problem instances. When considering $P^*$ in the upper plot of Fig.~\ref{fig:mks-results}, it is apparent that increasing $p$ consistently improves solution quality for IF+XY. At $p=1$, IF+XY performs similarly to the no-slack TAE, which outperforms the slack TAE in 16 of the 20 considered instances. IF+XY at $p=5,10$ exhibit drastically higher $P^*$ values, especially at $p=10$, where $P^* > 1\%$ for all instances.

The performance difference becomes even more prominent when examining $P_{90}$ results: IF+XY at $p =10$ maintains $P_{90} > 10\%$ (approaching $100\%$ for most cases), whereas TAE drops below $P_{90} < 1\%$ for larger problems.

The problem scenarios comprise two categories: Single-Knapsack instances (indices $<$10), where IF+XY reduces to IF since the reformulation \eqref{eq:mks-reformulated} is unnecessary, and complete MKS instances (indices $>$10) where the $XY$ search-space reduction becomes apparent. Overall, this comparison demonstrates that our constrained architecture IF+XY offers superior performance for this industry-relevant COP.

\subsection{Prosumer Problem}
\begin{table}
\setlength{\tabcolsep}{4.5pt}
\footnotesize
\centering
\caption{PP instances sorted by number of binary variables ($N$) and qubit requirements of IF+XY and QUBO}
\label{tab:problem-instances} 
\begin{tabular}{l|rrrl|rr}
\hline
Problem & $N$ & $m$ & $n$ & $d_i$ & {\scriptsize IF+XY} & {\scriptsize QUBO} \\ 
\hline
\hline
\texttt{Comb1\_0} & 6 & 5 & 2 & 3, 3 & 15 & 14 \\
\texttt{Comb0\_0} & 7 & 5 & 2 & 3, 4 & 22 & 17 \\
\texttt{Comb2\_0} & 7 & 5 & 2 & 4, 3 & 23 & 19 \\
\texttt{Comb1\_2} & 10 & 7 & 2 & 5, 5 & 25 & 22 \\
\texttt{Comb0\_2} & 11 & 7 & 2 & 5, 6 & 32 & 25 \\
\texttt{Comb2\_2} & 11 & 7 & 2 & 6, 5 & 35 & 29 \\
\texttt{Comb3\_0} & 13 & 5 & 3 & 5, 4, 4 & 32 & 23 \\
\texttt{Comb5\_0} & 13 & 6 & 3 & 4, 4, 5 & 37 & 31 \\
\texttt{Comb4\_0} & 14 & 6 & 3 & 3, 6, 5 & 38 & 32 \\
\texttt{Comb3\_2} & 19 & 7 & 3 & 7, 6, 6 & 46 & 33 \\
\texttt{Comb5\_2} & 19 & 8 & 3 & 6, 6, 7 & 51 & 43 \\
\texttt{Comb4\_2} & 20 & 8 & 3 & 5, 8, 7 & 52 & 44 \\
\texttt{Comb6\_0} & 26 & 8 & 4 & 6, 6, 6, 8 & 58 & 50 \\
\texttt{Comb7\_0} & 33 & 10 & 4 & 8, 9, 8, 8 & 73 & 63 \\
\texttt{Comb6\_2} & 34 & 10 & 4 & 8, 8, 8, 10 & 74 & 64 \\
\texttt{Comb8\_0} & 40 & 10 & 5 & 8, 8, 8, 8, 8 & 87 & 70 \\
\texttt{Comb7\_2} & 41 & 12 & 4 & 10, 11, 10, 10 & 89 & 77 \\
\texttt{Comb8\_2} & 50 & 12 & 5 & 10, 10, 10, 10, 10 & 107 & 86 \\
\texttt{Comb9\_0} & 51 & 10 & 6 & 8, 9, 8, 8, 9, 9 & 100 & 81 \\
\texttt{Comb10\_0} & 56 & 11 & 6 & 9, 9, 9, 9, 10, 10 & 110 & 89 \\
\texttt{Comb9\_2} & 63 & 12 & 6 & 10, 11, 10, 10, 11, 11 & 122 & 99 \\
\texttt{Comb10\_2} & 68 & 13 & 6 & 11, 11, 11, 11, 12, 12 & 132 & 107 \\
\texttt{Comb11\_0} & 74 & 12 & 7 & 10, 11, 10, 11, 9, 12, 11 & 144 & 110 \\
\texttt{Comb11\_2} & 88 & 14 & 7 & 12, 13, 12, 13, 11, 14, 13 & 170 & 130 \\
\hline
\end{tabular}
\vspace{-1em}
\end{table}

\subsubsection{Problem Instances}
The problem instances for the prosumer problem are generated such that the Knapsack constraints are binding, i.e., for every constraint, a solution candidate exists in the search space that violates it. Each instance combines $n$ pre-defined load profiles and a discrete time horizon $m$ with varying electricity price profiles. An overview of the problem parameters is given by Tab.~\ref{tab:problem-instances}, e.g., \texttt{Comb1\_0} tries scheduling two 2-hour loads inside 5 hours. The \texttt{\_2} suffix refers to the same instance as \texttt{\_0}, but enlarged by two time steps. Each problem consists of $N=\sum_id_i$ binary variables, where $d_i$ refers to a selection of variables that is one-hot constrained. For each problem ID in Tab.~\ref{tab:problem-instances}, we generate four noisy price patterns describing their general trend: increasing, decreasing, up-quadratic, and down-quadratic. All generated instances are available online, see Sec.~\ref{sec:data_avail}.

For the penalty value $\eta$, we empirically determined that setting it to the optimal objective value $f(x^*)$ returns the best results. While this approach is typically impractical in real-world applications---as the optimal value remains unknown before solving the problem---it enables a rigorous comparative evaluation against baseline QUBO methods by ensuring that we avoid inappropriate penalty values that might undermine the validity of our comparison.

\subsubsection{Solution Quality and Optimal Probability}

\begin{figure}
    \centering
    \includegraphics[width=\linewidth, trim={0 0.75cm 0 0},clip]{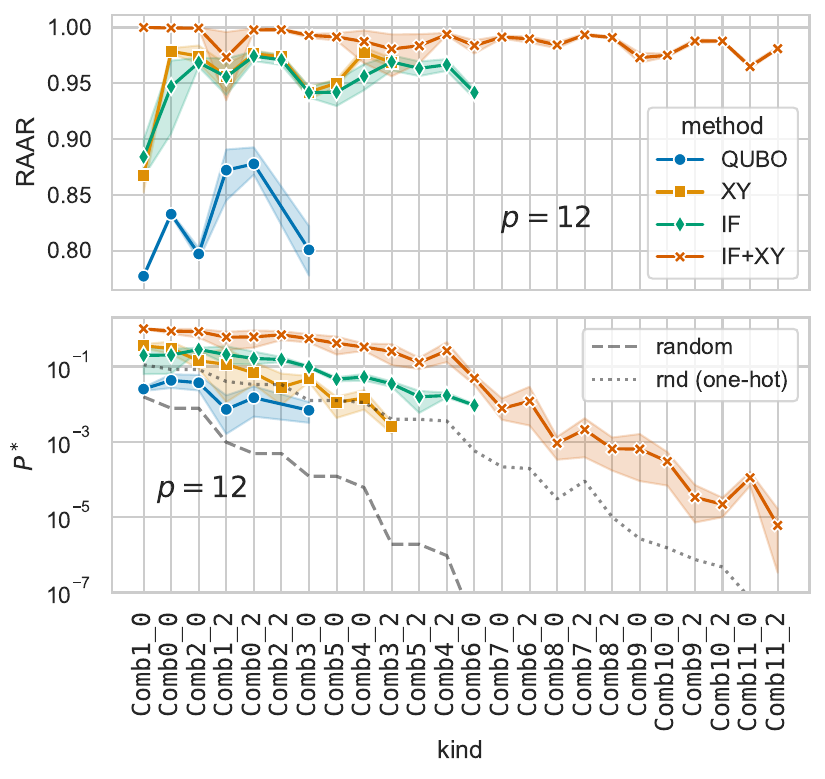}%
    \caption{Results showing the RAAR (upper panel) and $P^*$ (lower panel) for the four investigated pipelines at maximum QAOA layer depth $p=12$. The dashed and dotted lines in the lower plot mark probability when random sampling in the $N$ bits or the one-hot constrained subspace, respectively.}
    \label{fig:pp-raar}
    \vspace{-1em}
\end{figure}

The upper panel of Fig.~\ref{fig:pp-raar} shows the RAAR as a measure of solution quality at $p=12$. Notably, not all instances can be simulated with all pipelines, as explained in Sec.~\ref{sec:sim}, giving us the most data points for IF+XY and the least for QUBO. All results show strong performance with RAAR consistently above 0.75, i.e., considerably better than random sampling. In particular, IF+XY shows the best RAAR, remaining above 0.95 across all instances.

The lower panel exhibits a consistent pattern: IF+XY demonstrates the highest probability of measuring the optimal solution, followed by IF, XY, and QUBO. Nevertheless, the probability drops below $10^{-5}$ for IF+XY at \texttt{Comb11\_2} instance. Yet, a performance degradation is anticipated given the substantial size of this problem instance, which encompasses 88 binary variables (excluding ancillary qubits). The specialized simulation techniques introduced in Sec.~\ref{sec:sim} enable us to reach into that size domain previously inaccessible for ordinary state vector simulation.

All pipelines show a higher optimal probability than random sampling (dashed line). However, when considering random sampling within only the one-hot constrained subspace (dotted line), both the XY and QUBO approaches partly exhibit lower probability. Even though XY, by design, only samples in the one-hot-constrained subspace, the likelihood of sampling the perfect solution compared to random is reduced. This can be explained by a boost in the probability of good quality but not optimal solutions, since the RAAR is still reasonable.

\begin{figure*}
    \centering
    \subfloat[TTS$^*$ w.r.t.\ search space including linear fit of logarithmic qunatities\label{fig:pp-tts-fit}]{\includegraphics[width=0.5\linewidth]{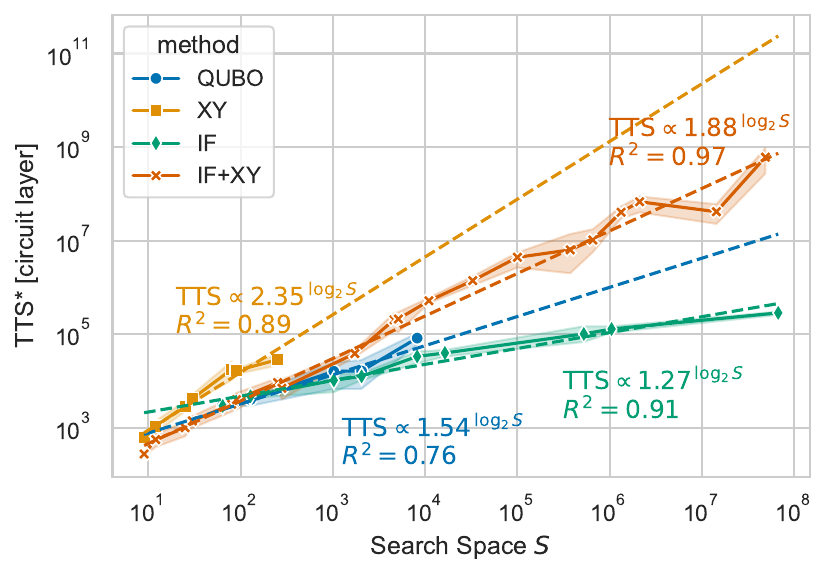}}%
    \subfloat[TTS$^*$ w.r.t.\ problem instance including extrapolation based on the fit\label{fig:pp-tts-extrapolation}]{\includegraphics[width=0.5\linewidth]{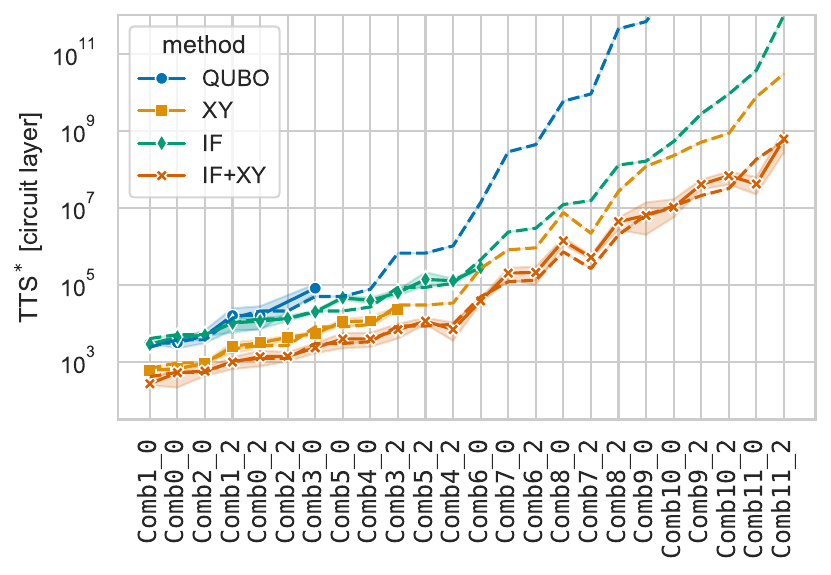}}
    \caption{Both panels of this figure show the TTS$^*$ of the four different pipelines. In panel (a), TTS$^*$ is plotted depending on the search space. The data points follow a power-law relationship, as indicated by their linear appearance on the log-log scale. Therefore, we fit linear trend lines into the plot, shown as dashed lines. Note that the slopes are not directly comparable due to different search spaces (One can compare IF+XY with XY, and IF with QUBO). Since the search space can be computed from the problem structure, we can extrapolate the TTS$^*$ results for the individual problem instances, as shown in panel (b).}
    \label{fig:pp-tts-full}
\end{figure*}

\subsubsection{Scaling of TTS}

Now, we focus on TTS$^*$ and its scaling characteristics. 
Fig.~\ref{fig:pp-tts-extrapolation} shows the TTS$^*$ concerning the previous PP instances. We first analyze only the measured data, postponing discussion of the dashed extrapolation lines. IF+XY consistently delivers the fastest overall TTS$^*$, even though the more complex circuit architecture. The higher $P^*$, as seen in Fig.~\ref{fig:pp-raar}, compensates for the higher circuit depth by reducing the required shots. Nevertheless, the differences between the pipelines have narrowed overall compared to the previous results.  The IF approach performs comparably to QUBO for smaller problem instances but exhibits more favorable scaling with problem size. While XY begins with lower TTS* values, it shows a steeper growth rate than IF. The IF+XY combination achieves both superior absolute performance and the most promising scaling behavior. Quantifying this advantage as a speedup factor $r = \frac{\text{TTS}^*_\text{QUBO}}{\text{TTS}^*_\text{IF+XY}}$, we observe a substantial average speedup of $r=87$ across the instances until \texttt{Comb3\_0}.

Based on these results, we aim to estimate the range of TTS$^*$ values to be expected for the complete dataset. However, the PP does not have a single problem size parameter; instead, when the number of time steps grows, the number of inequality constraints increases, and if the number of loads grows, the number of one-hot constraints increases. Of course, both constraints have some cross-correlation, i.e., even though the number of one-hot constraints does not increase with growing time steps, the number of binary variables within one constraint does.

To this end, we choose to correlate TTS$^*$ against the search space, which is the number of all possible candidate solutions of a method. Note that the search space differs from the state tensor's dimension since slack variables are discarded, leaving us with the same search space for QUBO and IF ($S = 2^N$), and XY and IF+XY ($S = \prod_i d_i$). The relation between $\log_2\text{TTS}^*$ and $\log_2 S$ is shown in Fig.~\ref{fig:pp-tts-fit}, alongside a fit for all methods. The regression lines show respectable $R^2$ values, but direct comparison of slopes across all methods is inappropriate due to their different search spaces. Only the pipelines with the same search space can be compared, showing that more constraint-preserving techniques help the scaling behavior by lowering the slope incline, e.g., the scaling of IF+XY is favorable compared to XY.

Since the search space can be directly computed for every PP instance, we extrapolate TTS$^*$ in Fig.~\ref{fig:pp-tts-extrapolation} (shown as the dashed lines). The QUBO TTS$^*$ appears to increase most drastically as the problem instances grow larger, though this projection should be interpreted cautiously, given the limited data points available for QUBO. XY and IF are expected to scale similarly with increasing problem sizes, but IF+XY maintains a performance advantage compared to them that potentially reaches more than two orders of magnitude approaching \texttt{Comb11\_2}.

\subsection{Discussion}\label{sec:discussion}

The presented empirical results show strong evidence that the suggested combined constraint pipeline IF+XY solves these constrained COPs most efficiently. Even though the results show substantial improvement compared to established methods, an open question still remains. Namely, the similar pattern of random sampling in the one-hot constrained feasible subspace and the IF+XY $P^*$ in Fig.~\ref{fig:pp-raar}, which is also underlined by the scaling base of 1.88 in Fig.~\ref{fig:pp-tts-fit} being close to 2. Does IF+XY, even with better sampling probability, scale similarly to random sampling? Of course, sub-exponential scaling is not expected for QAOA, but two possible answers exist: 

One reason is that we limit evaluation to $p=12$; deeper layers could likely lead to better TTS. However, data analysis shows that IF is the method that benefits the most from increased layers. For IF+XY, the ideal TTS is, on average, reached at $p=5.5$. Even though solution quality and $P^*$ grow with $p$, the growth rate is too small to offset the increased circuit depth. This can be explained by the optimization process that increases solution quality by boosting the amplitudes of high-quality non-optimal states. Examining the scaling behavior computed from $P_{90}$ could give another perspective.

The second reason is the performance of low-circuit-depth ring mixers, which may work well for small $d$ qudits but can deliver too little amplitude transfer for larger $d$, as we have in our instances (e.g., $d>10$). At least $p=d/2$ QAOA layers are required to theoretically guarantee amplitude transfer between all states.
To investigate whether this affects the performance, one needs to examine full $XY$-Mixers implemented through Trotterization, which may have prohibitive circuit layer depth for improving TTS.
Alternatively, Grover-Mixers with the $\ket{W}$-state preparation circuit could be studied to determine whether they enhance the scaling behavior.

%% file: sections/60_conclusion.tex
\section{Conclusion}\label{sec:conclusion}

In this work, we introduced a combined constraint architecture for QAOA that enforces one-hot constraints through $XY$-mixers and penalizes inequality constraints efficiently using the oracle-based IF method. Due to the circuit structure of QAOA, the restriction to the feasible subspace, and classically pre-computable IF, we employ several enhancements to accelerate and enable simulation of problem instances that would be otherwise impossible to simulate, e.g., up to 88 binary variables and 14 inequality constraints. The implementation relies on a combination of \texttt{custatevec} and \texttt{cutensor} methods from NVIDIA's \texttt{cuQuantum} SDK.

Benchmarking the combined and partial architecture against the default QUBO transformation has shown that IF+XY consistently outperforms the baseline variants in terms of RAAR (solution quality), $P^*$, and TTS. In the MKS experiments, we compared against the TAE from Ref.~\cite{hess2024} and obtained significantly better results, especially $P_{90}$ remained consistently above 10\%, while TAE drops below $1\%$.

For PP instances, we focus mainly on the holistic TTS$^*$ figure of merit and its scaling behavior. We fitted a linear regression between the logarithmic search space and logarithmic TTS$^*$ to extrapolate the expected runtime for the problem sizes and (partial) constrained QAOA architecture that cannot be simulated. IF+XY remained the fastest algorithm despite having the largest circuit overhead due to the mixer and oracle construction.

However,
the scaling of IF+XY is still remarkably close to the random sampling from all one-hot constrained states. Namely, $\text{TTS}^*\propto1.88^{\log_2 S}$ compared to $\propto S$, which might counterbalance the algorithm improvements in large-scale problems.

Nevertheless, there are several routes to continue work from here. For once, we suspect that the $XY$-ring mixer may be a dominant limitation here. Even though the circuit required for implementing this mixer is shallow, each mixer layer's narrow causal distance (1-2 neighbors) can be obtrusive. Therefore, revisiting these benchmarking instances with more sophisticated architectures, like a full $XY$-mixer through Trotterization~\cite{wang2020} or Grover mixers~\cite{bartschi2020}, would gain more insights. We expect worse initial TTS$^*$ for small problems but improved TTS$^*$ at larger instances due to advantageous scaling.

Besides improving the performance of the current structure, extending the capabilities of this pipeline by supporting more constraints is another path forward. For instance, the set-packing constraints of the MKS can directly be enforced using Grover Mixers with fitting state preparation~\cite{bartschi2020}, similar to the construction of the $\ket{W}$-states.

This work has demonstrated successfully that our problem-centric workflow for generating QAOA circuits outperforms default QUBO methods for two Knapsack-based optimization problems. Further application to other industry-relevant optimization problems has to show its general applicability. Also, since current hardware is progressing steadily, experiments on real quantum devices must be conducted to examine the NISQ-era performance of this more complex architecture.

\section{Data Availability}
\label{sec:data_avail}

The problem instances and entire simulation data are available in the following repository: \href{https://github.com/aqarios/multi-constrained-qaoa-data/tree/fixed_data}{github.com/aqarios/multi-constrained-qaoa-data}.

\section*{Acknowledgements}
We thank Naeimeh Mosheni, Kumar Ghosh from E.ON, and Jonas Nüßlein from LMU for their helpful discussions. This work was supported by the German Federal Ministry of Education and Research under the funding program “Förderprogramm Quantentechnologien – von den Grundlagen zum Markt” (funding program quantum technologies — from basic research to market), project Q-Grid (13N16177) and QuCUN (13N16199). J.S. acknowledges support from the German Federal Ministry for Economic Affairs and Energy through the funding program "QuantumComputing – Applications for the industry" based on the allowance "Development of digital technologies" (contract number: 01MQ22008A).